\documentclass[twocolumn]{aa}
\usepackage{graphicx}
\usepackage{txfonts}

\def\ms{\,m\,s$^{-1}$}         
\def\kms{\,km\,s$^{-1}$}       
\def\ogly{Oglie}
\def\peg{Pegasid}

\begin{document}
   \title{The ''missing link'': 
   a 4-day period \\ transiting exoplanet around OGLE-TR-111\thanks{Based on observations 
	   collected with the FLAMES+UVES spectrograph at the VLT/UT2 Kueyen telescope 
	   (Paranal Observatory, ESO, Chile)}}


   \author{F. Pont\inst{1}, F. Bouchy\inst{2}, D. Queloz\inst{1}, N. C. Santos\inst{3}, C. Melo\inst{4}, 
           M. Mayor\inst{1},  \and S. Udry\inst{1}
          }

   \offprints{F. Pont\\
              \email{frederic.pont@obs.unige.ch}}

   \institute{Observatoire de Gen\`eve, 51 ch. des Maillettes, 1290 Sauverny, Switzerland
	  \and
              Laboratoire d'Astrophysique de Marseille, 
               Traverse du Siphon, 13013 Marseille, France
         \and
	     Lisbon Observatory, Tapada da Ajuda, 1349-018 Lisboa, Portugal
         \and
	      ESO, Casilla 19001, Santiago 19, Chile  
            }

   \date{Received ; accepted }

   \abstract{We report the discovery of a transiting hot Jupiter around OGLE-TR-111, from our radial velocity follow-up of OGLE transiting candidates in Carina. The planet has a mass of $0.53 \pm 0.11\ M_{\rm J}$ and a radius of $1.0^{+0.13}_{-0.06}\ R_{\rm J}$. Three transiting exoplanets have already been found among OGLE candidates, all with periods near 1.5 days. The planet presented here, with $P=4.0$ days, is the first exoplanet detected by transits with the characteristics of a "normal" hot Jupiter, as found in abundance by radial velocity surveys The radius of OGLE-TR-111b and the scarcity of hot Jupiters detected among OGLE transit candidates tend to indicate that the case of HD209458b, with a radius of 1.4 $R_{\rm J}$, is exceptional, with most hot Jupiters being smaller.

   }

   \authorrunning{F. Pont et al.}
   \titlerunning{Transiting exoplanet around OGLE-TR-111}

   \maketitle
 
\section{Introduction}

Photometric searches for transiting exoplanets are fast emerging as a worthy competitor of radial velocity surveys
to detect and characterize hot Jupiters.  About 120 exoplanet candidates have been discovered 
by radial velocity surveys since 1995, but because of the nature of the method,
only the orbital period and a lower limit on their mass is provided. The observation of planetary transits, together with 
radial velocity measurements, also yield the exact mass and the planetary radius, provinding
much tighter constraints for planet models.

The first exoplanet transit observed was that of HD209458b
(Charbonneau et al. 2000; Henry et al. 2000),  a planet already known from Doppler surveys (Mazeh et al. 2000).

Within the past year, three exoplanets have been discovered by their photometric transit (OGLE-TR-56, Konacki et al. 2003.; OGLE-TR-113 and OGLE-TR-132, Bouchy et al. 2004a), thanks to the OGLE planetary transit survey\footnote{The OGLE survey (Optical Gravitational Lensing Experiment) announced the 
detection of 137 short-period multi-transiting objects with transits shallower than 8 percents (Udalski et al. 2002ab, 2003).}. However, these three planets share a very unusual  characteristic compared to those discovered by Doppler surveys: they all have very short periods near 1.5 days,  much below the observed pile-up of periods in hot Jupiters at 3-4 days, even much lower than the hot Jupiter with the shortest period known, 2.5 days (Udry et al. 2003). 

Here we will call "\peg s" the hot Jupiters with $P=2.5\!-\!10$ days and dub "\ogly s" those with $P<2.5$ days, in recognition of the pivotal role of the OGLE survey in their discovery. The phrase "hot Jupiter" is used for all short-period gas giant planets.

The result of the OGLE survey posed two related problems of coherence: why are "\ogly s" absent from Doppler surveys, and where are the much more abundant \peg s in the transit surveys ? The radial velocity surveys show that \peg s in the solar neighbourhood are at least an order of magnitude more abundant than "\ogly s", and HD209458, with a radius of $\sim\!1.4\ R_J$, indicates that they can easily cause eclipses deep enough to be detected by the OGLE survey. Even accounting for the fact that the time sampling of the OGLE survey favours the detection of short-period transits, the absence of any \peg\ was difficult to explain. 

In this Letter we report the discovery of just such an object around OGLE-TR-111. Its characteristics provide some indication as to why the OGLE survey may have missed most hot Jupiter transits.

\section{Observations and reductions}

The spectroscopic observations were obtained during a 26-hour run spread
on 8 nights on FLAMES/VLT in March 2004 (program 72.C-0191).
FLAMES is a multi-fiber link which feeds into the spectrograph UVES up to 
7 targets on a field-of-view of 25 arcmin diameter, in addition to  
simultaneous ThAr calibration. The fiber link produces a stable 
illumination at the entrance of the spectrograph, and the 
ThAr calibration is used to track instrumental drift. Forty-five minutes on a 
17 magnitude star yield a signal-to-noise ratio of about 8 corresponding 
to a photon noise uncertainty of about 30 {\ms} on an unrotating K dwarf star.
In our programme, we have observed all the OGLE candidates in the Carina field compatible with a planetary companion (Pont et al., in preparation) as well as 18 candidates in the OGLE Galactic bulge field (Bouchy et al. 2004b).

The spectra from the FLAMES+UVES spectrograph were reduced 
and the radial velocities were computed following the procedure described 
in Bouchy et al. (2004a). However, a significant improvement was obtained 
by using a specific numerical correlation mask tailored to the spectral type 
of the target. Furthermore, the calculation of the radial velocity from the 
Cross-Correlation Function (CCF) was improved. We find that our new procedure 
significantly improves the radial velocity accuracy and pushes the systematics well 
below the 35 {\ms} quoted in Bouchy et al. (2004a).

\section{Results}

\subsection{Radial velocities}

\begin{figure}
\resizebox{8.5cm}{!}{\includegraphics{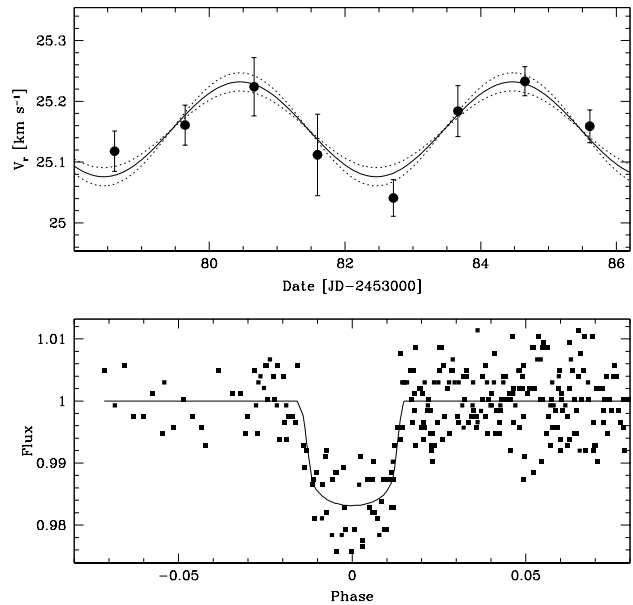}}
\caption{{\bf Top:} Radial velocity measurements for OGLE-TR-111, together with our orbital solution. {\bf Bottom:} Phase-folded normalized light curve and best-fit transit curve for OGLE-TR-111 (composed of nine partial individual transits).}
\label{doppler}
\end{figure}

Our radial velocity measurements and CCF parameters are listed in Table \ref{tablevr}. If the radial velocity
variations are caused by a transiting object, 
then the transit phase must correspond to the passage 
at center-of-mass velocity with decreasing velocity, and the period is fixed
by the periodicity of the transit signal. Given the short period, we assume that the orbit is circularized
with zero excentricity. We therefore fit on the data a sinusoid of  phase and period
fixed at the Udalski et al. (2002) values.
There are only two free parameters, the centre-of-mass velocity $V_0$ and the semi-amplitude $K$. We find $K=78 \pm 14$ \ms and $V_0=25.145 \pm 0.010$ \kms.  The radial velocity measurements and best-fit curve are displayed on Figure~\ref{doppler}.
The reduced $\chi^2$ is 4.2 for a constant velocity and 0.7 for a circular orbit.

\begin{table}
\caption{Radial velocity measurements (in the barycentric frame) and CCF parameters 
for OGLE-TR-111. }
\begin{tabular}{c c c c c c}\hline \hline
BJD & RV & depth & FWHM & S/N & $\sigma_{\rm RV}$  \\ 
{[$-2453000\,$d]} & [{\kms}] & [\%] & [{\kms}] & & [{\kms}] \\ \hline 
78.60419 &  25.118 & 31.27 & 9.2 & 7.4 & 0.033 \\
79.64235 &  25.161 & 30.75 & 9.0 & 7.3 & 0.033 \\
80.65957 &  25.224 & 28.18 & 9.0 & 5.5 & 0.048 \\
81.59492 &  25.112 & 25.46 & 9.4 & 4.5 & 0.067 \\
82.71279 &  25.041 & 31.09 & 9.1 & 8.0 & 0.030 \\
83.66468 &  25.184 & 29.14 & 9.0 & 6.2 & 0.042 \\
84.65149 &  25.233 & 33.57 & 9.0 & 9.5 & 0.024 \\
85.60720 &  25.159 & 32.84 & 9.1 & 8.5 & 0.027 \\ \hline
\end{tabular}
\label{tablevr}
\end{table}

 Leaving the period as a free parameters in the velocity fit does not cause any significant change, and no period other than 4 days produces an adequate orbit fit.

\subsection{Spectral classification}

On the summed spectra, the intensity and equivalent width of some spectral lines were analyzed 
to estimate the temperature, gravity and metallicity of the primary in the manner described in 
Santos et al. (2003). The resulting spectroscopic parameters are given in Table~\ref{tablespectro}. 
Although the low signal-to-noise ratio of the spectra prevents a very precise determination, the data 
indicates that OGLE-TR-111 is a late-G to early-K dwarf of solar or higher metallicity.

The full width at half-maximum (FWHM) of the CCF shows that the rotation velocity is lower than 5 {\kms},
indicating that the target is not tidally locked with its companion.

\subsection{Light curve analysis and physical parameters}

OGLE-TR-111 was measured 1176 times by the OGLE survey during one season in 2002. Nine individual transits were covered, with a depth of 1.9 percent and a periodicity of 4.01610 days (Udalski et al. 2003).

We fitted an analytical transit light curve to the photometric data by non-linear least-squares in the manner 
described in Bouchy et al. (2004b), to constrain the radius ratio, the
sum of masses, the primary radius and the impact parameter.

These constraints were then combined with the spectroscopic parameters, assuming that
OGLE-TR-111 is a normal dwarf star situated within the stellar evolution tracks
of Girardi et al. (2002), with the radial velocity orbit and the assumption of 
a circular Keplerian orbit seen almost edge-on. All the constraints were combined 
by $\chi^2$ minimisation in the way also described in details in Bouchy et al. (2004b). The resulting
values for the mass and radius of the star and planet are given in Table~\ref{tablephys}.
Note that because of the geometry of the problem, the uncertainties on several parameters are not symmetrical.
We find a mass of $0.53 \pm 0.11\ M_J$ and a radius of $1.00+^{+0.13}_{-0.06} \ R_J$ for OGLE-TR-111b.


\subsection{Triple-system blend scenarios}

\begin{figure}
\resizebox{8.5cm}{!}{\includegraphics{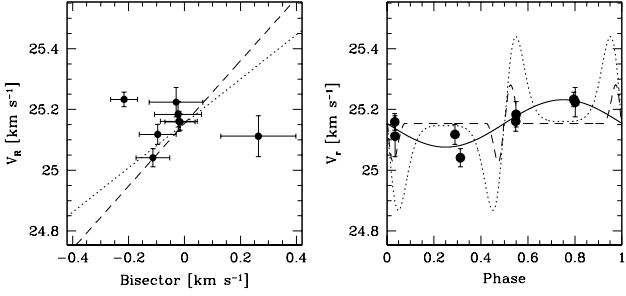}}
\caption{{\bf Top:} Bisector span [V$_{top}$ - V$_{bottom}$] of the three targets as a function of radial velocity. Uncertainties correspond to twice 
the radial velocity uncertainties. {\bf Bottom: }  Phased radial velocity data with the best-fit planet orbit as a solid line. In both plots, the prediction of two blend scenarios are shown. Dotted: $M_p=0.7 M_\odot$,   $M_s=0.2 M_\odot$, dashed :  $M_p=0.3 M_\odot$,   $M_s=0.2 M_\odot$ }
\label{bisspan}
\end{figure}

In certain circonstances, a triple system, or a single star with a background unresolved eclipsing binary, can mimick  a planet transit and produce phased velocity variations. In such cases, however, the CCF bisector is expected to vary. In order to examine the possibility that the radial velocity variations be due to a blend scenario, we computed the line bisectors as described by 
Santos et al. (\cite{santos}). Figure \ref{bisspan} shows that 
there is no significant correlation of the line asymmetries with phase. 

As a further check of blend scenarios, the cross-correlation function was computed with different 
masks (G2, K0 and K5 spectra) without significant change in the  amplitude of the orbital signals.
Blends of different spectral types are expected to produce velocity signals varying with the correlation mask.

More contraints can be put on a blend scenario for OGLE-TR-111 in the case of a physical triple system (by far the most probable case of false positive in the OGLE Carina field). Let us call "target", "primary" and "secondary" the components of such a triple system. The target is the star dominating the light and spectrum, the primary and secondary are the components of an associated close binary system. In the case of OGLE-TR-111, the only free parameter in a triple scenario is the mass of the primary. The mass and radius of the target are constrained by the spectroscopy, the masses, radii and luminosity of the three component are linked by main-sequence relations for low-mass stars, and the radius of the secondary is fixed by the transit depth. The radial velocity variation of the primary is governed by Kepler's law, the rotation velocity of the primary and secondary are fixed by tidal synchronisation.

The spectroscopy shows that the target is a low-mass star ($M\sim 0.8 \pm 0.1 M_\odot$). Using $R\!\!\sim\!\! M$ and $L\!\!\sim\!\! 10^{4 M} $ (approximations of the Baraffe et al. 1998 models for low-mass stars), the primary would have to be lighter than about 0.7 $M_\odot$ ($L_p/L_t <~15$ \%), otherwise its spectrum would cause a clearly visible signal in the CCF. It would have to be heavier than $\sim 0.3 M_\odot$ to contribute enough light to cause the transit signal in the combined photometry ($L_p/L_t\!\!>\!\! 2$ \%). For a triple system the transit depth is $d \simeq  (\frac{R_s}{R_p}) ^2 \frac{1}{1+L_t/L_p}$. With $K\! \sim \!P^{-1/3}\,m\,(m+M)^{-2/3}$, it turns out that all configurations imply a velocity semi-amplitude of the primary larger than 25 \kms. Because the target has a narrow CCF, the CCF signal of the primary would be separated from the CCF of the target during most of the phase. Two typical exemples from synthetic CCF simulations are shown on Figure~\ref{bisspan} for illustration. Even if the period is left as a free parameter, no scenario can reproduce the observed velocity variation and the absence of bisector variation. 

Other blend scenarios with a background binary or an equal-mass binary of double period all involve heavier stars, and therefore higher radial velocity amplitudes, making it even harder to reproduce the observed signal.

Therefore  no simple three-star blend scenario can be made compatible with the spectroscopic data for OGLE-TR-111.

\section{Discussion and Conclusion}

\begin{table}
\caption{Parameters for the  star OGLE-TR-111 and its planetary companion. }
\begin{tabular}{l l } \hline \hline
 & \\
Period [days]& 4.01610 [fixed]\\
Transit epoch [JD-2452000]&  330.44867 [fixed] \\
Radius ratio & $0.120 \pm 0.006$\\
Impact parameter & 0 - 0.68 \\
Inclination angle [$^o$] & 86.5 - 90 \\
Radial velocity semi-amplitude [\ms] & 78 $\pm$ 14\\
Systemic velocity [\kms] & 25.154 $\pm$ 0.010\\
O-C residuals [\ms] & 24\\
  & \\

Temperature  [K] & 5070 $\pm$ 400 \\
{[Fe/H]} & 0.12 $\pm$ 0.28\\
$\log g$ & 4.8 $\pm$ 1.0 \\
Star mass [$M_\odot$]&  $0.82^{+0.15}_{-0.02}$\\
Star radius [$R_\odot$]& $0.85^{+0.10}_{-0.03}$ \\
  & \\

Orbital semi-major axis [AU]& 0.047 $\pm$ 0.001\\
Orbital excentricity  & 0 [fixed] \\
Planet mass [$M_J$]& $0.53 \pm 0.11$\\
Planet radius [$R_J$]& 1.00$^{+0.13}_{-0.06}$ \\
Planet density [$g\,cm^{-3}$]& $0.61^{+0.39}_{-0.26}$ \\ \hline
\end{tabular}
\label{tablephys}
\label{tablespectro}
\end{table}

\begin{figure}
\resizebox{8.5cm}{!}{\includegraphics{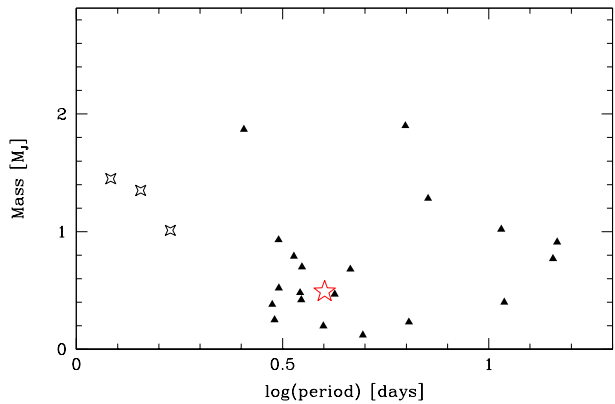}}
\caption{The period-mass relation for hot Jupiters from radial velocity (triangles) and transit (stars) surveys (in the first case, minimum masses only). The large star symbol is OGLE-TR-111b.  Contrarily to the three other OGLE transiting planets, OGLE-TR-111b is a "normal" hot Jupiter in terms of mass and period.}
\label{periodmass}
\end{figure}

OGLE-TR-111b is a typical \peg\ in terms of both period and mass (see Fig.~\ref{periodmass}). It corresponds to the pile-up of periods near 3-4 days observed in Doppler exoplanet surveys (Udry, Mayor \& Santos 2003), and its small mass is near the median of \peg\ masses, that are found to be much lighter on average than longer-period gas giants (Zucker \& Mazeh 2002). 

OGLE-TR-111b is only the second normal hot Jupiter for which a secure determination of radius has been obtained, and it has a markedly lower radius than the other case, HD209458 ($R\sim\!\! 1.4\ R_J$). OGLE-TR-111b is also clearly different from the three very short-period planets detected earlier in the OGLE survey (see Introduction), all three having masses superior to 1~$M_J$.

In some way, OGLE-TR-111b provides the "missing link" between planets from transit and radial velocity surveys. Both the absence of "normal" hot Jupiters and the abundance of $P\sim 1.5$ days planets in the OGLE survey were difficult to reconcile with the period distribution of the two dozens hot Jupiters known from radial velocity surveys. OGLE-TR-111b offers two crucial elements in the resolution of this mystery: 
First, its detection shows that \peg s are present in the OGLE transit survey.
Second, its radius shows that  HD209458 cannot be taken as a prototype of \peg s, therefore that many of them may have smaller radii, making their detection by transit much more difficult (the transit depth depending on the square of the planet radius).

In the context of the statistical interpretation of the detection of OGLE-TR-111b, it is worth noting that its period is very nearly  an integer number of days. This is probably no coincidence. Given the time sampling of ground-based transit survey, with a dominant 1-day frequency, resonant periods near multiples of 1-day can be detected even below the nominal detection threshold.  It is significant that, even with such a small-radius primary, this planet would probably not have been detected without its resonant period, which goes some way to explain the low detection rate of hot Jupiters by the OGLE survey.

Our result thus has sobering implications for the expected output of other ground-based exoplanet transit searches. If hot Jupiters can have radii similar to Jupiter, the accuracy and time coverage of the OGLE survey is sufficient to detect them only for a $R<R_\odot$ host star and with a resonant period. This implies that unless surveys have a better time  or volume coverage and/or better accuracy than the OGLE survey, they cannot be expected to detect \peg s in significant number.



The comparison of HD209458b and OGLE-TR-111b is compatible with simple expectations about the effect of stellar illumination on the structure of hot Jupiters: a gas giant close to a warm stars can have a large radius, low density and high evaporation rate, while around a cooler star it may stand nearer to the density of an isolated planet (HD209458 is $\sim\!\! 1000$ K hotter than OGLE-TR-111).  The discovery of more transiting planets, and  age information on the host stars, are obviously needed for a more detailed understanding.


\begin{figure}
\resizebox{8.5cm}{!}{\includegraphics{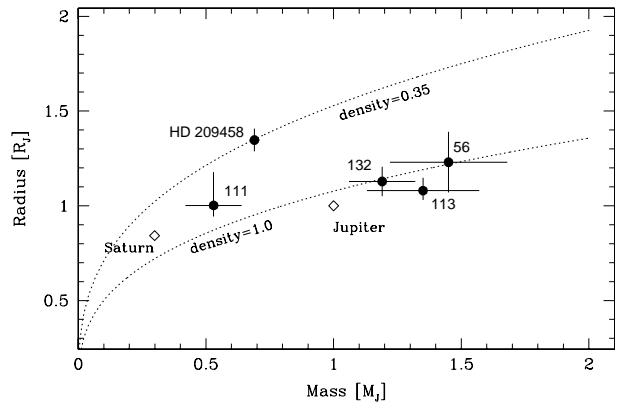}}
\caption{The mass-radius relation for transiting exoplanets. Jupiter and Saturn are indicated for comparison, as well as the loci of isodensities at 1.0 $g\,cm^{-3}$ and 0.35 $g\,cm^{-3}$. Data from Brown et al. 2001 (HD 290458), Konacki et al. 2003 (OGLE-TR-56), Bouchy et al. 2004a (OGLE-TR-113), Moutou et al. 2004 (OGLE-TR-132) and this Letter (OGLE-TR-111). }
\label{massradius}
\end{figure}

\begin{acknowledgements}
FP thanks the Swiss PRODEX fund for support. Support from Funda\c{c}\~ao para a Ci\^encia e Tecnologia (Portugal) to N.C.S. in the form of a scholarship is gratefully acknowledged.
\end{acknowledgements}

\end{document}